# Resonance modes in stereometamaterial of square split ring resonators connected by sharing the gap

Sheng Lei Wang,[1] Jun Jun Xiao,[1,*] Qiang Zhang,[1] and Xiao Ming Zhang[1]

[1]*College of Electronic and Information Engineering, Shenzhen Graduate School, Harbin Institute of Technology, Shenzhen 518055, Guangdong, China*
[*]*eiexiao@hitsz.edu.cn*

**Abstract:** Stereometamaerials can fully utilize the 3D degrees of freedom to exploit the coupling and hybridization between multiple split ring resonators (SRRs), enabling more extraordinary resonances and properties over their planar counterparts. Here we propose and numerically study a kind of structure based on connected SRRs sharing their gap in a rotational fashion. It is shown that there are three typical resonance modes in such cage-like SRR (C-SRR) stereometamaterial in the communication and near infrared range. In the order of increasing energy, these modes can be essentially ascribed to magnetic torodial dipole, magnetic dipole, and a mixture of electric-dipole and magnetic toroidal dipole. We show that the latter two are derived from the second-order mode in the corresponding individual SRR, while the first one from the fundamental one. The highest energy mode remains relatively "dark" in an individual C-SRR due to the high-order feature and the rotational symmetry. However, they are all easily excitable in a C-SRR array, offering multiband filtering functionality.

## 1. Introduction

Artificial subwavelength metamaterials can realize a lot of peculiar electromagnetic phenomena that are not in existence or much weaker in natural material. Split ring resonator (SRR) that can be treated as a magnetic dipole atom is widely used in metamaterial design, supporting magnetic resonance with magnetic or electric excitation [1,2]. As the basic and fundamental meta-atoms for constructing metamaterial, SRRs are organized in different ways, most frequently and particularly in two-dimensional spatial arrangement. Many planar combinations of SRRs have been proposed and fascinating phenomena have been realized from the microwave and infrared to near-infrared and even the visible [3-5]. Environmental or structural design of SRR can result in unusual resonances, such as dark mode [5] and trapped mode [6]. Moreover, Fan et al experimentally observe toroidal dipolar response and Fano resonance in a planar metamaterial [7].

Metamaterials in planar or coplanar form are of many advantages, particularly in easy fabrication. However, they cannot fully utilize the three dimensional degrees of freedom to exploit the coupling between SRRs and the hybridization of their resonances which are supposed to be more intriguing. Stereometamaterial, as first coined and studied by Liu et al [8], has been gaining arising attention over the past years. When SRRs are allowed to be arranged in three-dimensional space in deliberate configuration, they can offer more combination possibilities and the fundamental resonance mode can couple in a more comprehensive way. This could certainly lead to more interesting controls over the electromagnetic field in subwavelength scale. For example, Tsai et al [9] present a two plasmonic metamaterial composed of four SRRs supporting toroidal dipole response at optical frequencies. And they apply the high Q-factor toroidal resonance of the plasmonic toroidal metamaterial to lasing spaser [10]. Dong et al [11] designed a feasible nanostructured metamaterial that also support toroidal dipolar response in the optical regime by asymmetric double-bar. The double-bar structure can be regarded as a special SRR in generating magnetic resonance. Fedotov et al [12] experimentally demonstrate a new kind of toroidal metamaterials that implement non-trivial non-radiating charge-current excitation based on interfering electric and toroidal dipoles which is first proposed by Afannasiev and Stepanovsky [13]. The toroidal dipolar moment violates the symmetries of both space-inversion and time-reversal simultaneously and exhibits sub-radiating property, due to weak coupling to free space. It is essentially characterized by vortex distributions of head-to-tail magnetic dipoles and can be produced by currents flowing on the surface of a torus along its meridians [14]. Although acknowledged in atoms [15], molecules [16] and ferroelectric [17] structures, toroidal moment is hard to detect because being overwhelmed by electric and magnetic multipoles. Stereometamaterial, however, offers a rather feasible way to construct notable toroidal dipolar mode in electrodynamics [18].

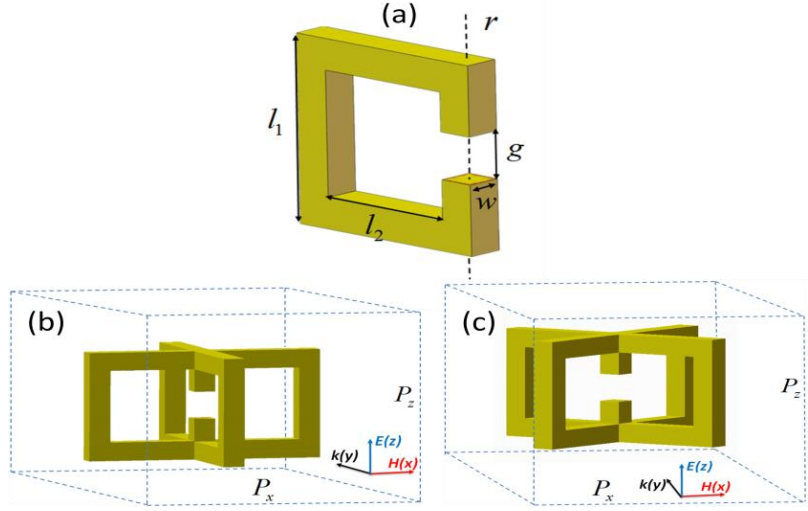

Fig. 1. Schematic diagram of the stereometamaterial. (a) Feature sizes of the constitution SRR where $l_1 = 300$ nm, $l_2 = 200$ nm, $w = 50$ nm, $g = 75$ nm and the dashed line labels the rotation axis (z-axis) for generating the C-SRR stereometamaterials. (b) and (c) unit cell of the C-SRR array with periods $P_x$ and $P_z = 600$ nm, respectively. (b) The incident light is incoming with **k**-vector parallel to two of the SRRs and perpendicular to the other two SRRs. (c) The C-SRR is rotated by 45° degree with respect to the normal direction of the array plane.

In this paper, we propose and study a kind of stereometamaterial structure consisting of multiple SRRs that are connected in a rotational fashion and share the split gap. It is shown that the structure sustains hybridized resonance modes resulting from the fundamental and the second-order modes in the corresponding individual SRR. When making them in periodic array, a higher-order mode due to formation of stronger electric dipole and toroidal dipole resonance would emerge (or become easily excitable). The three resonances of the proposed stereomatamaterial respectively lie in the communication, near-infrared and visible band. We show that the visible resonance is more sensitive to the periodicity and the arrangement, corresponding to high symmetry and high-order phase relationships in the constitute SRRs.

## 2. Stereometamaterial structure and numerical analysis

Figure 1 illustrates the geometry of the proposed stereometamaterial. The basic constitute unit is a square SRR as shown in Fig. 1(a). The design parameters can be found in the figure caption. The proposed stereometamaterial is constucted by duplicated SRRs that are rotated by 90°, 180°, and 270° degree with respect to the gap-bearing side, as shown in Fig. 1(b). Namely, four SRRs are connected by sharing the same split arm and gap. The structure can therefore be regarded as two orthogonally connecting rectangular sheets with "H"-shaped aperture, and resembles a cage. For short and simplicity let us call it as C-SRR. We study stereometamaterials of C-SRRs placed in a two-dimensional array with periodicity $P_x$ and $P_z$. The array is excited by $z$ polarized light at normal incidence with one sheet in the $xoz$ plane and the other parallel to the $k$ direction [Fig. 1(b)]. We also consider the array with the entire C-SRR unit rotated by 45° degree [Fig. 1(c)]. Numerical calculations are carried out by solving the three-dimensional Maxwell equations with the finite element method (COMSOL Multiphysics). The material of the SRRs is assumed to be gold whose permittivity is taken from Johnson and Christy [19] and the background medium is set as air. Perfectly matched boundary condition and periodic boundary condition are employed to obtain the scattering cross section and the transmission spectrum in the simulation, respectively.

In order to analyze and quantify which multipole component contributes most to the resonance peaks, the radiated powers of the electric and magnetic multipoles and toroidal dipole are calculated by the induced volume current density **j** in the C-SRR [9, 12, 20], e.g., electric dipole moment:

$$\mathbf{P} = \frac{1}{i\omega}\int \mathbf{j}\, d^3 r \quad (1)$$

Magnetic dipole moment:

$$\mathbf{M} = \frac{1}{2c}\int (\mathbf{r}\times\mathbf{j})\, d^3 r \quad (2)$$

Magnetic toroidal dipole moment:

$$\mathbf{T} = \frac{1}{10c}\int \left[(\mathbf{r}\cdot\mathbf{j})\mathbf{r} - 2r^2 \mathbf{j}\right] d^3 r \quad (3)$$

Electric quadrupole moment:

$$Q^{(e)}_{\alpha\beta} = \frac{1}{i2\omega}\int \left[r_\alpha j_\beta + r_\beta j_\alpha - \frac{2}{3}(r\cdot j)\delta_{\alpha\beta}\right] d^3 r \quad (4)$$

Magnetic quadrupole moment:

$$Q^{(m)}_{\alpha\beta} = \frac{1}{3c}\int \left[(r\times j)_\alpha r_\beta + (r\times j)_\beta r_\alpha\right] d^3 r \quad (5)$$

The radiated power of all the multipole moments sums over their contributions as:

$$I_{C-SRR} = \frac{2\omega^4}{3c^3}|\mathbf{P}|^2 + \frac{2\omega^4}{3c^3}|\mathbf{M}|^2 + \frac{4\omega^5}{3c^4}\operatorname{Im}(\mathbf{P}^*\cdot\mathbf{T}) + \frac{2\omega^6}{3c^5}|\mathbf{T}|^2 + \frac{\omega^6}{5c^5}\sum\left|Q^{(e)}_{\alpha\beta}\right|^2 + \frac{\omega^6}{40c^5}\sum\left|Q^{(m)}_{\alpha\beta}\right|^2 \ldots \quad (6)$$

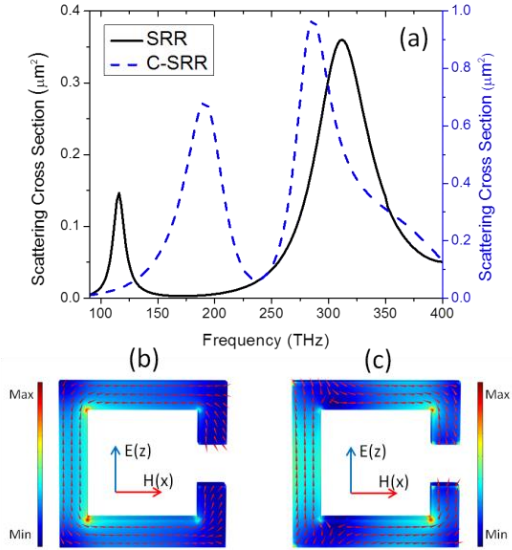

Fig. 2. (a) Scattering cross section of single SRR and an individual cage-SRR (C-SRR). (b) and (c) the surface current corresponding to the resonance peaks of single SRR. Background color is for the current intensity.

Note that the above formulas are in CGS unit and $c$ is the speed of light in vacuum, $\alpha,\beta = x, y, z$ the Cartesian coordinates. In view of the polarization and the symmetry of the C-SRR, we may focus on the dominant component from each moment. For example,

$IP_z = 2\omega^4 P_z^2 / 3c^3$ means the z-component ($P_z$) radiation power from the electric dipole **P** [20]. Similar rules apply to the other higher-order moments shown in Eqs. (1)-(6).

Before going directly to the two-dimensional array, we first examine single SRR and single C-SRR. Figure 2(a) shows the resonance spectra of an individual SRR (solid curve) and an isolated C-SRR (dashed curve). The SRR has the fundamental resonance $|\omega_1\rangle$ at $f = 116$ THz and its second-order resonance $|\omega_2\rangle$ at $f = 312$ THz, both excitable with incoming light polarized along the gap direction [see Figs. 2(b) and 2(c)]. The induced current for the $|\omega_1\rangle$ mode forms a loop that starts and ends at the gap boundaries. However, for the $|\omega_2\rangle$ mode, currents in the upper and lower edges are anti-parallel, but with no continuity at the vertical edge. There are two nodes near the corners [see Fig. 2(c)]. The C-SRR in this case is illuminated by a plane wave as shown in Fig. 1(b) and the scattering cross section (dashed line) also has two obvious peaks, one at $f = 192$ THz and the other at $f = 286$ THz. Notice, however, that there is an additional small side-peak at around $f = 370$ THz, at the shoulder of the $f = 286$ THz peak. We will later show that both of them originate from the $|\omega_2\rangle$ mode in the corresponding SRR. Since the individual SRR resonance can be tuned by the geometry (e.g., $l_1$, $l_2$, $g$ and $w$) from the microwave to the near-IR [3]. The C-SRR resonances can be flexibly controlled by these geometry parameters as well. More importantly, C-SRR with hetero-structured SRRs (e.g., making one SRR with shorter or longer $l_2$) can offer more resonance peaks (figure not shown here), in view of the broken $C_4$ rotational symmetry. Here we would like to focus on C-SRR with homo-structured SRRs.

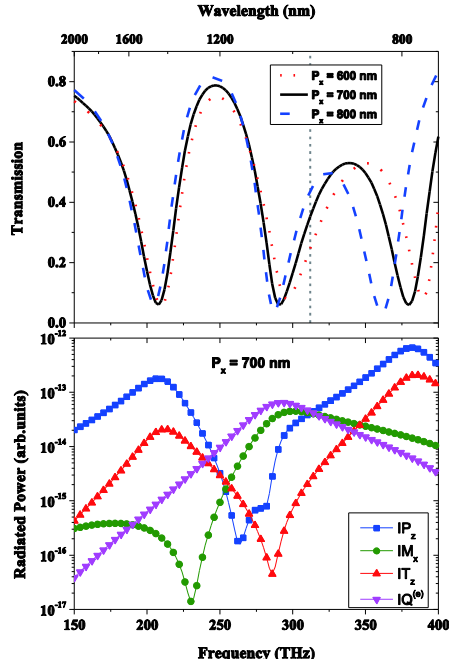

Fig. 3. (a) Simulated transmission of the two dimensional arrays shown in Fig. 1(b), with periodic boundary conditions for $P_x = 600$ nm, 700 nm, and 800 nm. The vertical dot line represents the frequency of $|\omega_2\rangle$ mode in single SRR. (b) Dispersion of radiated power from the various multipole moments when $P_x = 700$ nm.

Figure 3 shows the calculated results for C-SRR array. Firstly, we study the situation as shown in Fig. 1(b). The transmission spectrum is shown in Fig. 3(a) in which three resonances are clearly seen at $f = 208$ THz, 290 THz, and 380 THz for $P_x = 700$ nm (black curve). For easy explanation, we use the notation of $|\omega_{c1}\rangle$, $|\omega_{c2}\rangle$, and $|\omega_{c3}\rangle$ for them, respectively. The resonance frequency of $|\omega_2\rangle$ mode in the SRR is marked by the vertical dashed line whereas $|\omega_1\rangle$ is far below this range. When the periodicity is reduced ($P_x = 600$ nm) or increased ($P_x = 800$ nm), the positions of the first two resonance peaks remain nearly unchanged, but the highest frequency one is more apparently shifted to $f = 362$ THz and $f = 390$ THz, respectively. Figure 3(b) shows the radiated powers defined by Eqs. (1)-(6) as a function of the frequency, for the case of $P_x = 700$ nm and $P_z = 600$ nm. It is seen that for the resonance modes $|\omega_{c1}\rangle$ and $|\omega_{c3}\rangle$, the electric dipole moment $P_z$ and the magnetic toroidal dipole moment $T_z$ provide dominating contributions which are orders of magnitude larger than those from the other moments. The resonance mode $|\omega_{c2}\rangle$ is mainly contributed by the magnetic dipole moment $M_x$ and an electric quadrupole moment $Q^{(e)}$.

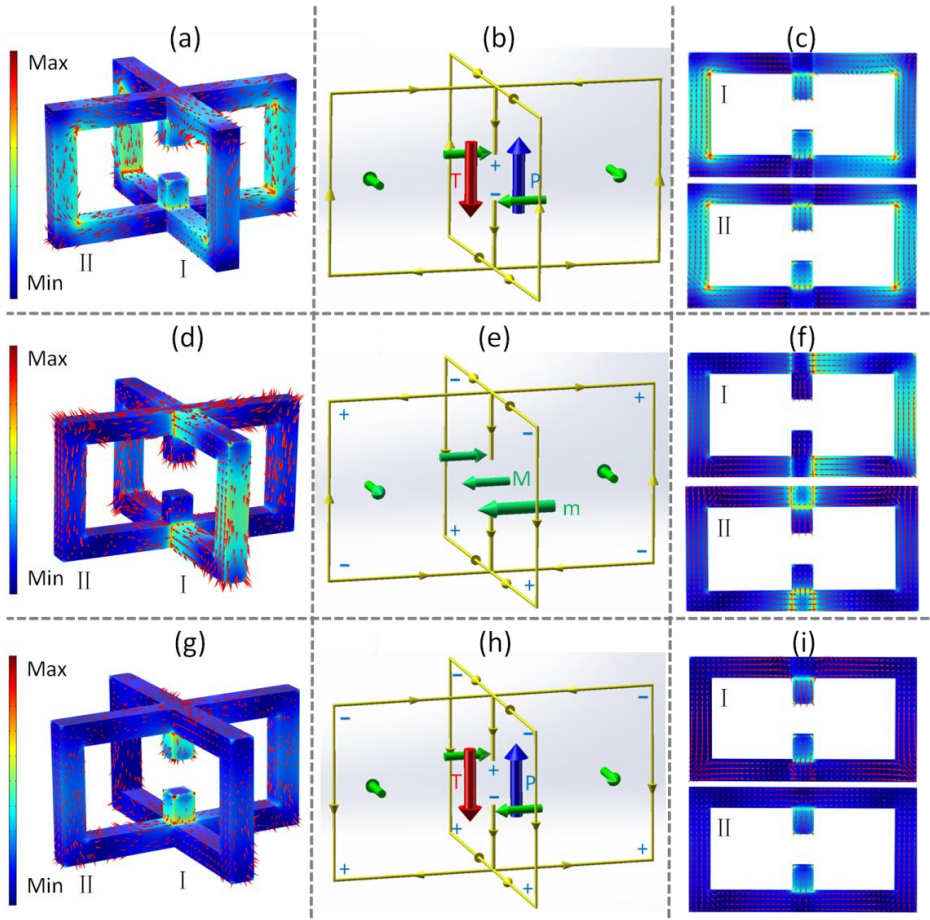

Fig. 4. Surface current density (arrows) maps corresponding to the three transmission dips in Fig. 3. (a), (d) and (g) for the C-SRR unit cell. (b), (e) and (h) schematic diagram of the induced current and effective electric dipole moment (blue arrow), magnetic dipole moment (green arrows), toroidal dipole moment (red arrow), and accumulated charges ("+" and "-"). (c), (f) and (i) currents on sheets I and II of the C-SRR in Figs. 4(a), 4(d) and 4(g), correspondingly.

Figure 4 plots both visually and schematically the current distribution for the three resonance modes observed in Fig. 3(a). Figures 4(a)-4(c) are for $|\omega_{c1}\rangle$ at $f = 208$ THz, Figs. 4(d)-4(f) for $|\omega_{c2}\rangle$ at $f = 290$ THz, and Figs. 4(g)-4(i) are for $|\omega_{c3}\rangle$ at $f = 380$ THz. Figure 4(a) shows that the current loop in each arm of the C-SRR is of the same pattern as in Fig. 2(b). Figures 4(a) and 4(c) show clearly that the current in the four arms flow from the same side of gap to the other uniformly, i.e., all of them loops in phase and are of $C_4$ rotational symmetry with respect to the $z$ axis. This develops four head-to-tail magnetic dipole moments in $xy$ plane, and collectively generates a remarkable toroidal dipole moment $\mathbf{T}$. Opposite charges are accumulated across the SRRs gap which yield an electric dipole moment $\mathbf{P}$ pointing reversely to the toroidal dipole moment [see Fig. 4(b)]. Notice that in this case, the current amplitudes are comparable in the four SRRs (shown in color scale).

In Figs. 4(d) and 4(f), it is seen that the distribution of the induced current in each SRR arms corresponds to that in Fig. 2(c). In this case, the current in the vertical edges is in opposite direction, i.e., out-of-phase to the ones in the other two edges, preventing a complete loop in the SRR. The upper and lower edges in each SRR support anti-parallel currents which generate four magnetic dipoles that appear cancelling each other, as schematically shown in Fig. 4(e). However, due to the retardation effect, the front SRR has much stronger current than those in the back SRR and in sheet **II**. As a result, there is a residual magnetic dipole moment $M_x$. Simultaneously, at the corners of the SRR arm, positive or negative charges are accumulated, forming an electric quadrupole moment $Q^{(e)}$. In this regard, $C_4$ rotational symmetry is broken in the $|\omega_{c2}\rangle$ mode. These results are in accordance to the radiation spectra in Fig. 3(b).

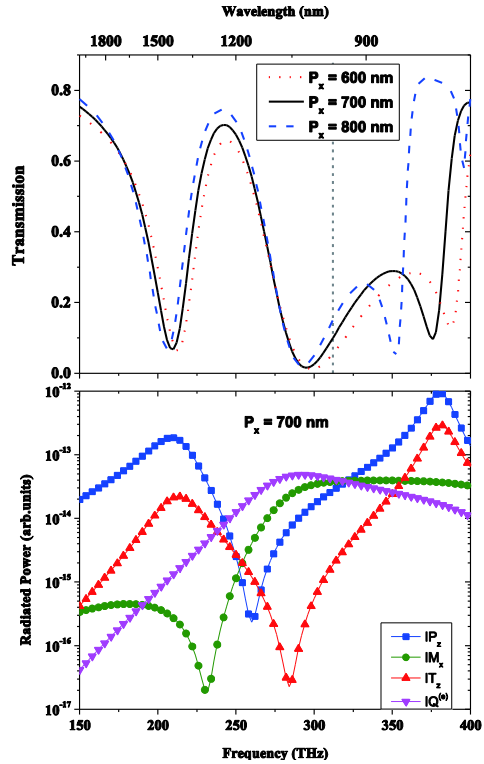

Fig. 5. (a) Simulated transmission of the two dimensional arrays shown in the situation of Fig. 1(c). The gray dot line represents the second-order mode $|\omega_2\rangle$ of single SRR. (b) Dispersion of the radiated powers of various multipole moments for $P_x = 700$ nm.

Figures 4(g)-4(i) show that for the period-sensitive mode $|\omega_{c3}\rangle$ at $f = 380$ THz, the currents on each SRR arm are also of the same pattern as that in Fig. 2(c). However, they are in different configurations to the $|\omega_{c2}\rangle$ mode. More specifically, the oscillating phase is $\pi$-inverted in the two SRRs that are perpendicular to the $k$ direction (i.e., in sheet **II**), with respect to that in sheet **I**. In this way, the current distributions on the four SRRs are again of $C_4$ rotational symmetry. The currents in the upper and lower edges flow in opposite direction and generate head-to-tail magnetic dipoles in a circle [see Fig. 4(h)], effectively captured by the magnetic toroidal dipole moment **T**. Notice that there are also collective positive and negative charges at the corners that form four electric dipoles which point in the opposite direction with respect to the big electric dipole over the SRRs gap. The strength of these corner-charge induced electric dipoles would be much smaller. These make this mode relatively dark and not easy to be excited in the isolated case. In this perspective, the $|\omega_{c3}\rangle$

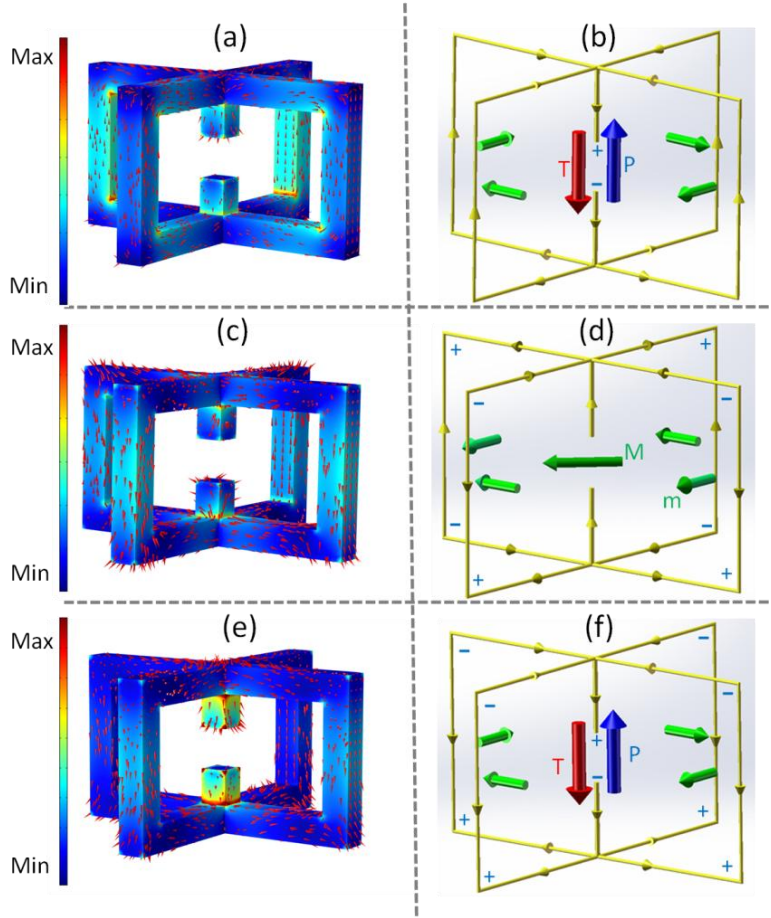

Fig. 6. Same as Fig. 4, while the C-SRR unit is rotated by 45° degree with respect to the normal direction of the array plane [see Fig. 1(c)].

($|\omega_{c2}\rangle$) mode can be regarded as a symmetric (anti-symmetric) hybridization of the $|\omega_2\rangle$ mode in a rotational fashion

Finally, in view of the hybridization mode symmetry, we would like to examine the situation where the incident light does not see any difference of sheets **I** and **II** in the C-SRR. The case is represented by Fig.1(c). Figures 5(a) and 5(b) show the calculated transmission spectra and the radiation power spectra, respectively. As expected, the $|\omega_{c1}\rangle$ and $|\omega_{c2}\rangle$ modes remain relatively stable at the same frequencies at $f = 210$ THz and $f = 296$ THz, while the $|\omega_{c3}\rangle$ mode shifts from $f = 352$ THz to 376 THz, and 388 THz for $P_x = 600$ nm, 700 nm, and 800 nm, respectively. Quite similarly, the three resonances can be regarded as from a magnetic toroidal dipole **T**, a magnetic dipole **M**, and a mixture of electric dipole and toroidal dipole [see Fig. 5(b) which shows the relative contribution from the multipole moments]. Figures 6(a) and 6(b) show the details of resonance mode $|\omega_{c1}\rangle$. It is seen that four magnetic dipoles loop in a circle, representing a typical magnetic torodial dipole. Figures 6(e) and 6(f) show that the resonance mode $|\omega_{c3}\rangle$ has higher-order resonance features and can be regarded as hybridization between $|\omega_2\rangle$ mode of the individual SRR in a $C_4$ rotational symmetry. In this configuration, we can see that the $|\omega_{c1}\rangle$ and $|\omega_{c3}\rangle$ resonance modes are basically of the same origin as discussed in Fig. 4. However, the total magnetic dipole moment and the total quadrupole moment of $|\omega_{c2}\rangle$ mode are produced directly by the nearly equal current loops of the constituent SRRs [see Figs. 6(c) and 6(d)]. This is different to the situation as shown in Figs. 4(d)-4(f). Notice that the transmission goes to nearly zero at this point of $f = 296$ THz [see Fig. 5(a)]. Clearly, the rotated C-SRR structure becomes more easily excitable by incoming plane wave. This can bring more convenience in possible application of sensing.

## 3. Conclusion

In summary, we show that hybridized resonance modes can be engineered in a three-dimensional cage-like stereometamaterial made of square SRRs that sharing a gap. The resonances derive from the fundamental and the second-order modes in individual SRR. The three dimensional configuration and the close connection make it possible to generate magnetic toroidal moment in either lower energy or higher energy, enabled by rotational symmetric current loops in the SRRs. Break the rotational symmetry of the SRR current loops results in a magnetic dipole at intermediate energy. Our result may provide a platform for study of toroidal moments with higher-order magnetic resonances. We believe it could be useful in engineering photon emission and in sensing application.

### Acknowledgments

This work was supported by NSFC (No. 11274083), and the Shenzhen Municipal Science and Technology Plan (Grant Nos. KQCX20120801093710373, JCYJ20120613114137248, and 2011PTZZ048). JJX is also supported by the Natural Scientific Research Innovation Foundation in Harbin Institute of Technology (No. HIT.NSFIR.2010131). We acknowledge helps from the Key Lab of Terminals of IoT and the National Supercomputer Shenzhen Center.